\documentclass[12pt, a4paper]{article}
\usepackage{amsfonts,amssymb,amsmath,amsthm,graphicx}

\begin{document}

\title{Casimir energy of two plates inside a cylinder.
\thanks{A talk given at the 14th International Seminar on High Energy
Physics QUARKS-2006, St.Petersburg, Russia, 19-25 May , 2006.} }

\author{Valery N.Marachevsky \thanks{email: maraval@mail.ru} \\
{\it V. A. Fock Institute of Physics, St. Petersburg
State University,}\\
{\it 198504 St. Petersburg, Russia} }



\maketitle

\begin{abstract}
The new exact formulas for the attractive Casimir force acting on
each of the two perfectly conducting plates moving freely inside
an infinite perfectly conducting cylinder with the same cross
section are derived at zero and finite temperatures by making use
of zeta function technique. The short and long distance behaviour
of the plates' free energy is investigated.
\end{abstract}

\maketitle

\section{Introduction}

Recently a new geometry in the Casimir effect\cite{Casimir}, a
piston geometry, has been introduced  in a $2D$ Dirichlet model
\cite{Cavalcanti}. Generally the piston is located in a
semi-infinite cylinder closed at its head. The piston is
perpendicular to the walls of the cylinder and can move freely
inside it. The cross sections of the piston and cylinder coincide.
Physically this means that the approximation is valid when the
distance between the piston and the walls of a cylinder is small
in comparison with the piston size.

In paper \cite{Jaffe2} a perfectly conducting square piston at
zero temperature was investigated in $3D$ model in the
electromagnetic and scalar case. The exact formula (Eq.(6) in
\cite{Jaffe2}) for the force on a piston was written in the
electromagnetic case. Also the limit of short distances was found
for arbitrary cross sections (Eq.(7) in \cite{Jaffe2}). This
result was generalized in \cite{Marach1}. In paper \cite{Marach1}
the exact formula for the free energy of two perfectly conducting
plates of an arbitrary cross section inside the waveguide (or
infinite cylinder) with the same cross section was written
(Eq.(118) in \cite{Marach1} or (\ref{r21}) here). In the zero
temperature case and square cross section of the waveguide our
general result for the force (\ref{r30}) coincides with Eq.(6) in
\cite{Jaffe2}. It is interesting that our result resembles the
result for the interquark potential in a rigid string model
\cite{Nesterenko} .

A dilute circular piston and cylinder were studied perturbatively
in \cite{Barton}. In this case the force on two plates inside a
waveguide and the force in a piston geometry differ essentially.
The force in a piston geometry can even change sign in this
approximation for thin enough walls of the material. Other
examples of repulsive pistons were presented in \cite{Fulling}.

In Sec.2 we derive the new exact result (\ref{r21}) for the free
energy of two parallel plates inside an infinite cylinder by
making use of the zeta function technique \cite{Santangelo2,
Elizalde}. We consider a perfectly conducting case, the plates
move freely inside the cylinder with the same cross section, which
is arbitrary. The plates are perpendicular to the walls of the
cylinder. In Sec.3 we apply the heat kernel technique
\cite{Vassilevich, Gil, Marach1} to derive the leading short
distance behaviour of the free energy. In the short distance limit
we prove that there are no temperature corrections to the leading
terms obtained in \cite{Jaffe2} (Eq.(7) in \cite{Jaffe2}). The
long distance limit result (\ref{r24}) (the high temperature limit
result) is new.

We take $\hbar=c=1$.

\section{Derivation}

Our aim is to calculate the Casimir energy of interaction and the
force between the two parallel plates of an arbitrary cross
section inside an infinite cylinder of the same cross section (the
plates are perpendicular to the walls of the cylinder).

$TE$ and $TM$ eigenfrequencies of the perfectly conducting
cylindrical resonator with an arbitrary cross section can be
written as follows. For $TE$ modes ($E_z =0 $) inside the
perfectly conducting cylindrical resonator of the length $a$ with
an arbitrary cross section the magnetic field $B_z (x,y,z)$ and
eigenfrequencies $\omega_{ TE}$ are determined by:
\begin{align}
& B_{z} (x,y,z) = \sum_{i=1, n=1}^{+\infty} B_{in}
\sin \Bigl(\frac{\pi n z}{a}\Bigr) g_i (x,y) , \\
& \Delta^{(2)} g_i (x,y) = - \lambda_{i N}^2 g_i (x,y) \\
& \frac{\partial g_i (x,y)}{\partial n} \Bigl|_{\partial M} = 0 \\
& \omega_{TE}^2 =  \Bigl(\frac{\pi n}{a}\Bigr)^2 + \lambda_{i N}^2
, \:\: n=1 .. +\infty , i=1 .. +\infty.
\end{align}
The other components of the magnetic and electric fields can be
expressed via $B_z (x,y,z)$.

For the $TM$ modes ($B_z =0$) inside the perfectly conducting
cylindrical resonator of the length $a$ with an arbitrary cross
section the electric field $E_z (x,y,z)$ and eigenfrequencies
$\omega_{ TM}$ are determined by:
\begin{align}
& E_{z} (x,y,z) = \sum_{n=0, k=1}^{+\infty} E_{kn}
\cos \Bigl(\frac{\pi n z}{a}\Bigr) f_k(x,y) , \\
& \Delta^{(2)} f_k (x,y) = - \lambda_{k D}^2 f_k (x,y) \\
&  f_k (x,y) |_{\partial M} = 0 \\
& \omega_{TM}^2 =  \Bigl(\frac{\pi n}{a}\Bigr)^2 + \lambda_{k D}^2
, \:\: n=0 .. +\infty , k=1 .. +\infty
\end{align}

In  $\zeta$-function regularization scheme the Casimir energy is
defined as follows:
\begin{equation}
E = \frac{1}{2} \Bigl(\sum\omega_{TE}^{1-s} +
\sum\omega_{TM}^{1-s} \Bigr) \Bigl|_{s=0} , \label{a2}
\end{equation}
where the sum has to be calculated for large positive values of
$s$ and after that an analytical continuation to the value $s=0$
is performed.

Alternatively one can define the Casimir energy via a zero
temperature one loop effective action $W$ ($T_1$ is a time
interval here):
\begin{align}
&W = - E T_1  \\
&E = - \zeta^{\prime} (0) \label{a3} \\ &\zeta(s) =
\frac{1}{\Gamma(\frac{s}{2})} \int_{0}^{+\infty} dt\,
t^{\frac{s}{2} - 1} \sum_{\omega_{TE},\omega_{TM}}
\int_{-\infty}^{+\infty} \frac{d p}{2\pi} \exp \biggl( - t
\Bigl(\frac{a}{\pi}\Bigr)^2 \Bigl(\omega^2 +p^2\Bigr)\biggr)
\label{a1}
\end{align}
After integration over $p$ in (\ref{a1}) one can see that
definitions (\ref{a2}) and (\ref{a3}) coincide.

In every Casimir sum it is convenient to write:
\begin{equation}
\sum_{n=1}^{+\infty} \exp(- t n^2) = \frac{1}{2}
\sum_{n=-\infty}^{+\infty} \exp(- t n^2) - \frac{1}{2} =
\frac{1}{2} \, \theta_3 \Bigl(0, \frac{t}{\pi}\Bigr) - \frac{1}{2}
. \label{a4}
\end{equation}
For the first term on the right-hand side of (\ref{a4}) we use the
property of the theta function $\theta_3(0,x)$:
\begin{equation}
\theta_3 (0, x) = \frac{1}{\sqrt{x}} \, \theta_3
\Bigl(0,\frac{1}{x}\Bigr)
\end{equation}
and the value of the integral
\begin{equation}
\int_0^{+\infty} dt \, t^{\alpha-1} \exp \Bigl(-p\: t
-\frac{q}{t}\Bigr) = 2 \Bigl(\frac{q}{p}\Bigr)^{\frac{\alpha}{2}}
K_{\alpha} (2 \sqrt{p q} )
\end{equation}
for nonzero values of $n$ to rewrite the Neumann zeta function
$\zeta_N (s)$ (arising from TE modes) in the form:
\begin{multline}
\zeta_N (s) = \sum_{\lambda_{iN}} \int_{-\infty}^{+\infty}
\frac{dp}{2\pi} \biggl[
\frac{\sqrt{\pi}\:\Gamma\bigl((s-1)/2\bigr)}{2 \:\Gamma(s/2)}
\Bigl(\frac{a\sqrt{\lambda_{iN}^2 + p^2}}{\pi}\Bigr)^{1-s}
 + \\ + \sum_{n=1}^{+\infty} \frac{2\sqrt{\pi}}{\Gamma(s/2)}
\Bigl(\frac{\pi^2 n}{a \sqrt{\lambda_{iN}^2
+p^2}}\Bigr)^{\frac{s-1}{2}} K_{\frac{s-1}{2}} \Bigl(2an
\sqrt{\lambda_{iN}^2 +p^2}\Bigr) \biggr] -  \\ -
\sum_{\lambda_{iN}}\frac{\sqrt{\pi}\:\Gamma\bigl((s-1)/2\bigr)}{4
a \Gamma(s/2)} \Bigl(\frac{a \lambda_{iN}}{\pi}\Bigr)^{1-s}
\label{a5}
\end{multline}

The Neumann part of the Casimir energy is given by:
\begin{multline}
E_N = - \zeta_N^{\prime}(0) = \sum_{\lambda_{iN}}
\int_{-\infty}^{+\infty} \frac{d p}{2 \pi} \frac{1}{2} \ln
\Bigl(1-
\exp(-2 a \sqrt{\lambda_{iN}^2 +p^2}) \Bigr) +  \\
+ \frac{a}{2} \sum_{\lambda_{iN}} \int_{-\infty}^{+\infty} \frac{d
p}{2 \pi}  \Bigl( \lambda_{iN}^2 +p^2
\Bigr)^{\frac{1-s}{2}}\biggr|_{s=0} - \frac{1}{4}
\sum_{\lambda_{iN}} \lambda_{iN}^{1-s}\biggr|_{s=0} . \label{a6}
\end{multline}
Here we used $K_{-1/2} (x) =\sqrt{\pi/(2\,x)}\exp(-x)$.

The Dirichlet part of the Casimir energy (from TM modes) is
obtained by analogy:
\begin{multline}
E_D = \sum_{\lambda_{kD}} \int_{-\infty}^{+\infty} \frac{d p}{2
\pi} \frac{1}{2} \ln \Bigl(1-
\exp(-2 a \sqrt{\lambda_{kD}^2 +p^2}) \Bigr) +  \\
+ \frac{a}{2} \sum_{\lambda_{kD}} \int_{-\infty}^{+\infty} \frac{d
p}{2 \pi}  \Bigl( \lambda_{kD}^2 +p^2
\Bigr)^{\frac{1-s}{2}}\biggr|_{s=0} + \frac{1}{4}
\sum_{\lambda_{kD}} \lambda_{kD}^{1-s}\biggr|_{s=0}  .\label{a7}
\end{multline}

The electromagnetic Casimir energy of a perfectly conducting
resonator of the length $a$ and an arbitrary cross section is
given by the sum of (\ref{a6}) and (\ref{a7}) :
\begin{align}
E = &\sum_{\lambda_{iN}} \int_{-\infty}^{+\infty} \frac{d p}{2
\pi} \frac{1}{2} \ln \Bigl(1- \exp(-2 a \sqrt{\lambda_{iN}^2
+p^2}) \Bigr) + \label{a8} \\ + &\sum_{\lambda_{kD}}
\int_{-\infty}^{+\infty} \frac{d p}{2 \pi} \frac{1}{2} \ln
\Bigl(1- \exp(-2 a \sqrt{\lambda_{kD}^2
+p^2}) \Bigr) + \label{a9} \\
+ &\frac{a}{2} \sum_{\lambda_{iN}} \int_{-\infty}^{+\infty}
\frac{d p}{2 \pi}  \Bigl( \lambda_{iN}^2 +p^2
\Bigr)^{\frac{1-s}{2}}\biggr|_{s=0} + \label{a10} \\ +
&\frac{a}{2} \sum_{\lambda_{kD}} \int_{-\infty}^{+\infty} \frac{d
p}{2 \pi}
\Bigl( \lambda_{kD}^2 +p^2 \Bigr)^{\frac{1-s}{2}}\biggr|_{s=0} +
\label{a11}\\
+ &\frac{1}{4} \sum_{\lambda_{kD}} \lambda_{kD}^{1-s}\biggr|_{s=0}
- \frac{1}{4} \sum_{\lambda_{iN}} \lambda_{iN}^{1-s}\biggr|_{s=0}
. \label{a12}
\end{align}

The terms
\begin{align}
E_{waveguide} = &\frac{1}{2} \sum_{\lambda_{iN}}
\int_{-\infty}^{+\infty} \frac{d p}{2 \pi}  \Bigl( \lambda_{iN}^2
+p^2 \Bigr)^{\frac{1-s}{2}}\biggr|_{s=0} + \\ + &\frac{1}{2}
\sum_{\lambda_{kD}} \int_{-\infty}^{+\infty} \frac{d p}{2 \pi}
\Bigl( \lambda_{kD}^2 +p^2 \Bigr)^{\frac{1-s}{2}}\biggr|_{s=0}
\end{align}
yield the electromagnetic Casimir energy for a unit length of a
perfectly conducting infinite cylinder with the same cross section
as the resonator under consideration.

For the experimental check of the Casimir energy for the
rectangular cavity one should measure the force somehow. We think
about the following possibility: one should insert two parallel
perfectly conducting plates inside an infinite perfectly
conducting cylinder and measure the force acting on one of the
plates as it is being moved through the cylinder.  The distance
between the inserted plates is $a$.

To calculate the force on each plate the following gedanken
experiment is useful. Imagine that $4$ parallel plates are
inserted inside an infinite cylinder and then $2$ exterior plates
are moved to spatial infinity. This situation is exactly
equivalent to $3$ perfectly conducting cavities touching each
other. From the energy of this system one has to subtract the
Casimir energy of an infinite cylinder, only then do we obtain the
energy of interaction between the interior parallel plates, the
one that can be measured in the proposed experiment. Doing so we
obtain the attractive force on each interior plate inside the
cylinder:
\begin{align}
F (a) &= - \frac{\partial E_{arb}(a)}{\partial a}, \label{a13} \\
E_{arb} (a) &= \sum_{\omega_{wave}} \frac{1}{2} \ln (1-\exp(-2 a
\, \omega_{wave})), \label{r7}
\end{align}
the sum here is over all  $TE$ and $TM$ eigenfrequencies
$\omega_{wave}$ for the cylinder with an arbitrary cross section
and an infinite length. Thus it can be said that {\it the exchange
of photons with the eigenfrequencies of an infinite cylinder
between the inserted plates always yields the attractive force
between the plates}.

 For rectangular boxes it was generally believed \cite{Lukosz, Bordag1}
 that the repulsive contribution to the force acting on two parallel
opposite sides  of a {\it single} box (separated by a distance
$a$) and resulting here from (\ref{a10}-\ref{a11}) could be
measured in experiment. However, it is not possible to use the
expression (\ref{a8}-\ref{a12}) directly to calculate the force
since it includes the self-energy Casimir parts
(\ref{a10}-\ref{a11}) of the other sides of the resonator.
Nevertheless the expression (\ref{a8}-\ref{a12}) can be used to
derive a measurable force (\ref{a13}) between the freely moving
parallel plates inserted inside an infinite cylinder of the same
cross section as the plates.

To get the free energy $F_{arb} (a,\beta)$ for bosons at nonzero
temperatures ($\beta=1/T$) one has to make the substitutions:
\begin{align}
 p &\to p_m = \frac{2\pi m}{\beta} , \\
 \int_{-\infty}^{+\infty}  \frac{dp}{2\pi} &\to \frac{1}{\beta}
 \sum_{m=-\infty}^{+\infty}.
\end{align}
Thus the free energy describing the interaction of the two
parallel perfectly conducting plates inside an infinite perfectly
conducting cylinder of an arbitrary cross section has the form:
\begin{align}
F_{arb} (a,\beta) = &\frac{1}{\beta} \sum_{\lambda_{k D}}
\sum_{m=-\infty}^{+\infty} \: \frac{1}{2} \ln \Bigl(1-\exp
(-2a\sqrt{\lambda_{k D}^2 + p_m^2} ) \Bigr) + \nonumber \\ + &
\frac{1}{\beta} \sum_{\lambda_{i N}} \sum_{m=-\infty}^{+\infty} \:
\frac{1}{2} \ln \Bigl(1-\exp (-2a\sqrt{\lambda_{i N}^2 + p_m^2} )
\Bigr) , \label{r21}
\end{align}
where $\lambda^2_{k D}$ and $\lambda^2_{i Neum}$ are eigenvalues
of the two-dimensional Dirichlet and Neumann problems (a boundary
here coincides with the boundary of each plate inside the
cylinder):
\begin{align}
& \Delta^{(2)} f_k (x,y) = - \lambda_{k D}^2 f_k (x,y) \\
&  f_k (x,y) |_{\partial M} = 0 ,
\end{align}
\begin{align}
& \Delta^{(2)} g_i (x,y) = - \lambda_{i N}^2 g_i (x,y) \\
& \frac{\partial g_i (x,y)}{\partial n} \Bigl|_{\partial M} = 0 .
\end{align}

The attractive force between the plates inside an infinite
cylinder of the same cross section at nonzero temperatures is
given by:
\begin{align}
F (a, \beta) &= -\frac{\partial F_{arb} (a, \beta) }{\partial a} =
\nonumber
 \\ & - \frac{1}{\beta} \sum_{\omega_{TD}}
\frac{\omega_{TD}}{\exp(2a\omega_{TD}) - 1} - \frac{1}{\beta}
\sum_{\omega_{TN}} \frac{\omega_{TN}}{\exp(2a\omega_{TN}) - 1}.
 \label{r30}
\end{align}
Here $\omega_{TD} = \sqrt{p_m^2 + \lambda^2_{k D}} $ and
$\omega_{TN} = \sqrt{p_m^2 + \lambda^2_{i N}} $.

\section{Asymptotic cases}
It is convenient to apply the technique of the heat kernel to
obtain the short distance behaviour of the free energy
(\ref{r21}). It can be done by noting that if the heat kernel
expansion
\begin{equation}
\sum_{\lambda_i} e^{-t\lambda_i^2} \underset{t\to 0}{\sim}
\sum_{k=0}^{+\infty} t^{\frac{-n+k}{2}} c_k
\end{equation}
exists ($n$ is a dimension of the Riemannian space) then one can
write the expansion
\begin{equation}
\sum_{\lambda_i} e^{-\sqrt{t}\lambda_i} \underset{t\to 0}{\sim}
\sum_{k=0}^{n-1} \frac{2 \;\Gamma(n-k)}{\Gamma((n-k)/2)} \;\;
t^{\frac{-n+k}{2}} c_k  \label{r22}
\end{equation}
by making use of the analytical structure of the zeta function.
The strategy is the following: one expands the logarithms in the
formula (\ref{r21}) in series and then applies the expansion
(\ref{r22}) to each term.

For $a \ll \beta/(4\pi)$ one obtains from (\ref{r21}) and
(\ref{r22}) the leading terms for the free energy:
\begin{equation}
F_{arb}(a,\beta)|_{a \ll \beta/(4\pi)} =
-\frac{\zeta_R(4)}{8\pi^2}\frac{S}{a^3} - \frac{\zeta_R (2)}{4\pi
a} (1 - 2 \chi) + O(1) , \label{r23}
\end{equation}
where
\begin{equation}
\chi = \sum_i \frac{1}{24} \Bigl(\frac{\pi}{\alpha_i}-
\frac{\alpha_i}{\pi}\Bigr) + \sum_j \frac{1}{12\pi}
\int_{\gamma_j} L_{aa} (\gamma_j) d\gamma_j.
\end{equation}
Here $\alpha_i$ is the interior angle of each sharp corner and
$L_{aa} (\gamma_j)$ is the curvature of each smooth section
described by the curve $\gamma_j$. The force calculated from
(\ref{r23}) coincides with $F_C$ in \cite{Jaffe2}, (Eq.7).

In the opposite long distance limit $a \gg \beta/(4\pi)$ one has
to keep only $m=0$ term in $(\ref{r21})$. Thus the free energy of
the plates inside a cylinder in this limit (the high temperature
limit) is equal to:
\begin{align}
F_{arb}(a, \beta)|_{a \gg \beta/(4\pi)} =  &\frac{1}{2\beta}
\sum_{\lambda_{k D}}  \: \ln \Bigl(1-\exp (-2a\lambda_{k D} )
\Bigr) + \nonumber
\\ + & \frac{1}{2\beta} \sum_{\lambda_{i N}}
\: \ln \Bigl(1-\exp (-2a\lambda_{i N} ) \Bigr) \label{r24}
\end{align}
This result is new.

One can check that the limit $a \to 0$ in (\ref{r24}) immediately
yields the high temperature result for two parallel perfectly
conducting  plates separated by a distance $a$. One expands
logarithms in series and uses (\ref{r22}) and $c_{0D}=c_{0N}=
S/(4\pi)$ in two dimensions ($n=2$) to obtain:
\begin{multline}
F_{arb}(a, \beta)|_{a \gg \beta/(4\pi), a \to 0} = \\ =  -
\sum_{\lambda_{k D}} \frac{1}{2\beta} \sum_{n=1}^{+\infty}
\frac{\exp(-2an\lambda_{k D})}{n}\Bigl|_{a\to 0} -
\sum_{\lambda_{i N}} \frac{1}{2\beta} \sum_{n=1}^{+\infty}
\frac{\exp(-2an\lambda_{i N})}{n}\Bigr|_{a\to 0}
 = \\
= \sum_{n=1}^{+\infty} - \frac{1}{2\beta}
\frac{1}{n}\frac{1}{(2an)^2} 2 (c_{0D} + c_{0N}) =
-\frac{\zeta_R(3)}{\beta a^2} \frac{S}{8\pi}  ,
\end{multline}
which is a well known result \cite{Brevik, Sauer, Lifshitz}.

\section*{Acknowledgements}

  V.M. thanks D.V.Vassilevich, Yu.V.Novozhilov and V.Yu.Novozhilov
  for suggestions during the preparation of the paper.
  V.M. thanks R.L.Jaffe and M.P.Hertzberg
  for correspondence and discussions. This work has been
  supported by  grants RNP $2.1.1.1112$ and
  SS $.5538.2006.2$.

\end{document}